\documentclass[12pt,preprint]{aastex}

\def\simlt{\mathrel{\hbox{\rlap{\hbox{\lower4pt\hbox{$\sim$}}}\hbox{$<$}}}}
\def\simgt{\mathrel{\hbox{\rlap{\hbox{\lower4pt\hbox{$\sim$}}}\hbox{$>$}}}}

\def\lesssim{\mathbin{\lower 3pt\hbox 
      {$\rlap{\raise 5pt\hbox{$\char'074$}}\mathchar"7218$}}} 
\def\gtrsim{\mathbin{\lower 3pt\hbox
      {$\rlap{\raise 5pt\hbox{$\char'076$}}\mathchar"7218$}}}

\begin{document}

\title{The Magnetar Nature and the Outburst Mechanism of a Transient
Anomalous X-ray Pulsar}

\author{Tolga G\"uver$^{1}$, Feryal \"Ozel$^{2}$, Ersin
G\"o\u{g}\"u\c{s}$^{3}$ and Chryssa Kouveliotou$^{4}$}

\affil{$^{1}$Istanbul University, Science Faculty, Astronomy \& Space
Sciences Department, Beyaz\i t, Istanbul, 34119}
\affil{$^{2}$University of Arizona, Department of Physics, 1118 E. 4th
St., Tucson, AZ 85721}
\affil{$^{3}$Sabanc\i~University, Faculty of Engineering Natural Sciences, 
34956 Istanbul, Turkey}
\affil{$^{4}$NASA/MSFC, VP 62, 320 Sparkman Drive Huntsville, AL 35805, USA }

\begin{abstract}
Anomalous X-ray Pulsars (AXPs) belong to a class of neutron stars
believed to harbor the strongest magnetic fields in the universe, as
indicated by their energetic bursts and their rapid
spindowns. However, an unambiguous measurement of their surface field
strengths has not been made to date. It is also not known whether AXP
outbursts result from changes in the neutron star magnetic field or
crust properties.  
Here we report a spectroscopic measurement of the
surface magnetic field strength of an AXP, XTE J1810$-$197, and
solidify its magnetar nature. The field strength obtained from
detailed spectral analysis and modeling, B =
(2.72$\pm$0.03)$\times10^{14}$~G, is remarkably close to the value
inferred from the rate of spindown of this source and remains nearly
constant during numerous observations spanning over an order of
magnitude in source flux. The surface temperature, on the other hand,
declines steadily and dramatically following the 2003 outburst of this
source. Our findings demonstrate that heating occurs in the upper
neutron star crust during an outburst and sheds light on the transient
behaviour of AXPs.
\end{abstract}

\keywords{pulsars: general --- stars: individual (XTE J1810--197) --- 
stars: neutron --- X-rays: stars}
\section{Introduction}

The X-ray pulsar XTE J1810$-$197 was discovered (Ibrahim et al.\ 2004)
in 2003 when it suddenly brightened to more than 100 times its
quiescent value (Halpern \& Gotthelf 2005) during an outburst. The
source showed a steady decline of its X-ray flux thereafter,
accompanied by significant spectral changes (Gotthelf \& Halpern
2006). The 5.54~s period of the source, as well as the large period
derivative $\dot P \approx 10^{-11}$~s~s$^{-1}$ were established
(Ibrahim et al.\ 2004), confirming the source as an Anomalous X-ray
Pulsar (AXP). The detection of characteristic X-ray bursts (Woods et
al.\ 2005), similar to those seen in other AXPs (Gavriil, Kaspi, \&
Woods 2002, 2006), further strengthened this
classification.

XTE J1810$-$197 resides on one extreme of the diverse spectrum of
variability properties observed from AXPs. These span at least four
types of different activity in pulsed and persistent emission, ranging
from the short-lived energetic bursts to outbursts that are
characterized by sudden increases and subsequent long-timescale decays
in the persistent flux (Kaspi 2006). Moreover, XTE J1810$-$197 is also
unique in its unusually low quiescent flux levels that have been
determined from archival XTE and ROSAT data (Gotthelf \& Halpern
2006), almost two orders of magnitude fainter than the other known
AXPs, earning it the title of the transient AXP.

As with the X-ray spectra of the other known AXPs, the spectra of XTE
J1810$-197$ have so far been analyzed by fitting empirical functions
such as two blackbodies or a blackbody plus a power-law to the data
(Gotthelf \& Halpern 2005). Such analyses are typically used for
providing a rough estimate of the surface temperature of AXPs, even
though neutron star surfaces do not emit like blackbodies. The photon
energy-dependent radiation processes in their atmospheres strongly
distort the emission originating deep in the neutron stars away from a
blackbody spectrum (see, e.g., Shibanov et al.\ 1992). The strong
magnetic fields that the sources are thought to possess, based on the
rapid spindowns (Kouveliotou et al.\ 1998), also leave distinctive
imprints on the spectra both by altering the radiation processes in
their atmospheres (\"Ozel 2003) and by giving rise to moderate
scattering optical depths in the magnetospheres (Thompson, Lyutikov,
\& Kulkarni 2002).

In this Letter, we analyze the spectra of XTE J1810$-$197 for the
first time with a physical model that incorporates emission from the
magnetar surface and its reprocessing in the magnetosphere, as
described in G\"uver, \"Ozel, \& Lyutikov (2006). Unlike empirical
fits, we account for the known radiative processes that take place on
and around a magnetar for a range of magnetic field strengths and
surface temperatures and base our analysis on the resulting set of
models. The fits, therefore, can yield the physical parameters of the
source. Conversely, successfully reproducing in detail the spectral
characteristics of every epoch with a single physical model, while
keeping consistent values for these parameters, would indicate that
this model captures all the relevant physical effects that take place
on a magnetar and is thus a validation of the theoretical model. We
choose XTE J1810$-$197 a prime candidate for this detailed study,
because it has gone through extreme variations in its X-ray flux and
spectrum over its short lifetime and, thus, it is not a priori obvious
that such a wide range of spectra can be fit with a model that depends
only on four physical parameters.

We describe our physical model in Section 2 and present the data
analysis and results for XTE J1810$-$197 in Section 3. In Section 4,
we discuss the implications of the spectroscopically determined
magnetic field strength and conclude with a discussion of the
mechanism responsible for the outburst of the transient magnetar based
on the results of our analysis.

\section{The Surface Thermal Emission and Magnetospheric Scattering Model}

We base our spectral analysis on the physical model of magnetars that
we have developed (G\"uver et al.\ 2006), which for the first time
takes into account the relevant mechanisms that take place both on the
atmosphere and in the magnetosphere of a magnetar. The {\it Surface Thermal
Emission and Magnetospheric Scattering} model (STEMS) depends only on four
physical parameters that describe the surface magnetic field strength
and temperature of the neutron star, as well as the density and the
energetics of charges in its magnetosphere.

In our detailed calculations, we address the polarization-mode
dependent transport of radiation, treating absorption, emission, and
scattering processes that take place in the fully ionized plasmas of
hot ($\sim 0.1-0.6$~keV) magnetar atmospheres (\"Ozel 2003). The model
also incorporates the interaction of photons with the protons in the
plasma that gives rise to absorption features at the proton cyclotron
energy. Furthermore, we fully calculate the effects of vacuum
polarization resonance, which leads to an enhanced conversion between
photons of different polarization modes as they propagate outward
through the atmosphere. We have calculated spectra spanning the range
of surface magnetic field strengths $B = 5 \times 10^{13} - 5 \times
10^{15}$~G and surface temperatures $T = 0.1-0.6$~keV, in line with
the physical processes incorporated into the calculations.

In the stellar magnetospheres, we include a treatment of resonant
scattering (G\"uver, \"Ozel, \& Lyutikov 2006). The enhanced current
density in the magnetosphere of a magnetar significantly increases the
optical depth to electron scattering experienced by the outgoing
atmospheric photons (Thompson et al.\ 2002). The resulting
upscattering modifies both the high-energy continuum and the
equivalent widths of the proton cyclotron absorption features (Lyutikov
\& Gavriil, 2006; G\"uver et al., 2006).

Finally, because the surface photons originate in the strong
gravitational field of the neutron star, we follow the general
relativistic propagation of the photons to an observer on Earth. This
last step depends on the mass-to-radius ratio of the neutron star (for
which we assume a fixed fiducial value of $z= (1-2M/R)^{-1/2}-1 =
0.3$) and is necessary to make the physical models directly comparable
to the observations of AXPs. Here, $M$ and $R$ are the mass and radius
of the neutron star, respectively, given in gravitational units.

In order to compare observed X-ray spectra with the STEMS model, we
numerically calculated model X-ray spectra (in the 0.05 - 8.12 keV
range) and created a table model which can be used within XSPEC
(Arnaud 1996).

\section{Data Analysis and Results} 

XTE J1810$-$197 was observed for a total of 170 ks in seven pointings
between August 9, 2003 and March 12, 2006 with EPIC-PN onboard
XMM-Newton. During these observations, the unabsorbed $0.5-7$~keV flux
of the source varied from
57.96$\times~10^{-12}$~erg~s$^{-1}$~cm$^{-2}$ at its peak to
4.17$\times~10^{-12}$~erg~s$^{-1}$cm$^{-2}$ during the last
observation. Fast read out of EPIC-PN cameras and decreasing flux of
the source with time ensures that the spectra were not affected by
photon pile-up in any of the observations.

We calibrated all observations using the Science Analysis Software
(SAS, v. 7.0.0) and the latest (January 2007) available calibration
files. We eliminated the segments of data which were highly affected
by high-energy particle background. We grouped all the spectra before
background subtraction so that each spectral bin has at least 25
counts. We then fit the spectra, using XSPEC v11.3.2, with the STEMS
model we discussed above.  We allowed for interstellar extinction from
a cold medium with cosmic abundances.

Figure~1 shows the spectra observed in the seven different epochs, the
best fit models, and the residuals. The model describes in detail the
salient features of the spectra in the entire energy range. This is
especially remarkable because the significant evolution of the spectra
during the decay of the outburst can be fit with a single physical
model. The $\chi^2_{\nu}/$d.o.f. for the spectral fits, in order of
decreasing source flux are 1.07/732, 0.95/548, 1.11/820, 1.12/772,
1.21/653, 1.02/423, 1.02/302.  Even more compelling than the low
$\chi^2_{\nu}$ values is the flatness of residuals that demonstrate the
ability of the model to reproduce the observations without the need
for any additional ad hoc components such as a blackbody or a
power-law function.

In the analyses of all seven observations, we obtain a nearly
constant value of $N_{\rm H} = 0.63 \pm 0.08 \times10^{22}$~cm$^{-2}$
for the equivalent hydrogen column density responsible for the
interstellar extinction, even though we allow this parameter to vary
between observations. This value is also in agreement with an
independent study of this source (Gotthelf \& Halpern 2006). The
scattering optical depth $\tau$ and the velocity distribution $\beta$
of the electrons in the magnetosphere also remain fairly constant,
around values $\tau \approx 4.5 \pm 1.1$ and $\beta \approx 0.22 \pm
0.05$.  These results show that there is no significant variation in
the magnetosphere of the source during the window of the XMM-{\it
Newton} observations. If the onset of the outburst introduced any
magnetospheric changes, these must have stabilized within the months
between the peak of the outburst and the first XMM-{\it Newton}
observation.

\section{Discussion}

Our analysis of the 0.5-7~keV spectra of XTE J1810$-$197 with the
STEMS model allows for a tight and unique constraint of the magnetic
field strength of the source. The best-fit values for the surface
magnetic field and the surface temperature of the neutron star,
obtained from the detailed fits, as well as the 1- and 2-sigma
confidence limits are shown in Figure~2. The magnetic field strength
ranges from $B=(2.25 \pm 0.05) \times 10^{14}$~G in the earliest
observation to $B=(3.3 \pm 0.1) \times 10^{14}$~G in the last one,
while the temperature declines from $T=0.49 \pm 0.02$~keV to $T=0.22
\pm 0.03$~keV in the same interval. Note that for the last two
observations which have very low flux levels and poorer statistics,
the confidence contours were drawn by freezing the other model
parameters. For comparison, we also obtained the value of the magnetic
field strength from fitting all the data sets simultaneously, which
yields $(2.72 \pm 0.03) \times 10^{14}$~G.

The confidence contours show that the magnetic field can be tightly
constrained in each observation. This is because of the presence of
significant broad features in the magnetar spectra imparted by
weakened proton cyclotron lines and the vacuum polarization resonance
that have a strong dependence on the magnetic field strength. As we
previously discussed in G\"uver et al.\ (2006), these features allow
for a precise determination of the magnetic field strength from
continuum spectra. To demonstrate this effect, we plot in Figure~3 the
deviation of the model from the data obtained on March 18, 2005 (i.e.,
the longest observation) when the magnetic field strength is
artificially set to 8-sigma higher than the best-fit value, while all
the other parameters remain at their best-fit values (obtained before
modifying the field strength). The deviations are due to the broad
features that can be seen easily in the residuals. It is the detection
of these unique modifications in the spectra of XTE J1810$-$197 that
allows for the measurement of its surface magnetic field strength.

The contours shown in Figure 2 indicate that the surface magnetic
field strength and the surface temperature are not correlated. We find
that this is also true for all the other model parameters. As an
example, when we perform the above experiment where we set the
magnetic field strength to an articially higher value but, instead,
allow the other model parameters to change in the fit, we find a local
minimum with statistic value 805/653, which is higher than the global
minimum.

The measured magnetic field strength remains nearly constant during
the decline of the outburst. The slight evolution of the field
strength seen in Figure~2 likely arises from using phase-averaged
spectra in conjunction with the geometric simplicity of our physical
model that does not take into account variations of the field strength
on the neutron star surface. It may also be partially affected by the
spatial evolution of the hotspot that gave rise to the outburst on the
neutron star surface. Therefore, confirming or rejecting this trend
requires modeling that includes more details about the magnetic field
topology. The analysis of the pulse-phase resolved spectra (or
equivalently, the energy-dependent pulse profiles) of this source will
be reported elsewhere.

Our measurement is also in good agreement with the value of the
magnetic field inferred from the average spindown rate of the pulsar
using XMM-Newton and RXTE observations of the pulsar (Gotthelf \&
Halpern, 2006, Ibrahim et al.\ 2004), as well as from the
short-timescale measurements of the period derivative from radio
observations (Camilo et al.\ 2007).  Such an accord is unanticipated
given the numerous assumptions involved in inferring the magnetic
field strength with a vacuum dipole spindown formula (Spitkovsky
2006). Thus, our independent, spectroscopic
measurement of the magnetic field strength is a validation of the use
of the dipole spindown formula.

The surface temperature is also well constrained, as can be seen in
the tight confidence limits in Figure~2. The time arrow shows the
monotonic decline of the temperature during the sequence of the seven
XMM-{\it Newton} observations. The decay of the source flux during the
same time interval can be explained entirely by the cooling of the
neutron star crust, as described by the single temperature parameter,
without significant changes to the emitting area on the neutron star
surface. Assuming a distance of 3.3 kpc (Camilo et al. 2006), the
radius of this hot region remains approximately $3.7$~km, likely
corresponding to the area that is heated during the outburst. As
discussed earlier, the scattering optical depth $\tau$ and the
velocity distribution $\beta$ of electrons in the magnetosphere also
remain fairly constant, despite the changes in the hardness of the
spectra as the source cools. Indeed, the spectral changes are best
described by a change in the temperature alone, without accompanying
changes to any other parameter describing the neutron star surface or
its magnetosphere.

This physical model allows us to track the changes in the AXP during
its decline from its outburst and probe the mechanism that produces
the transient behavior. Suggested ideas for the observed flares and
outbursts for the AXPs and SGRs rely either on the injection of heat
deep in the crust or a sudden change in the topology of the field
lines in the magnetosphere. Our analysis, which disentangles the
contributions of the processes in the magnetosphere from those on the
stellar surface, shows that it is the release of heat in the crust,
and not changes in the magnetosphere, that is responsible for this long 
timescale AXP outburst.

We can identify the depth in the crust where the heat is released to
produce the outburst of XTE J1810$-$197 by considering the energetics
of the outburst and the measured evolution of the
temperature. Assuming that the heat is deposited over a surface area
$S$ at a depth $h$, where the particle density and temperature are
given by $N_{\rm d}$ and $T_{\rm d}$, respectively, we can calculate
the total energy of the outburst $E$ as $E \approx L \Delta t \approx
3 N_{\rm d} k_{\rm B} \Delta T_{\rm d} S h$. Here, $\Delta T_{\rm d}$
is the resulting increase in the temperature in the deep layer, which
is related to the change in the effective temperature by $\Delta
T_{\rm d} / T_{\rm d} \approx \Delta T_{\rm eff} / T_{\rm eff}$ by the
Eddington-Barbier relation. Assuming a distance of 3.3 kpc (Camilo et
al.\ 2006) and a timescale of $\Delta t \simeq 1$~yr, we estimate a
total energy of $10^{42}$~erg for the outburst using the typical
luminosity $L \simeq 3 \times 10^{34}$. Finally, we calculate the
particle density at a given depth using the detailed surface model of
the neutron star used in fitting the spectral data. Requiring that,
during the outburst, $\Delta T_{\rm eff}/T_{\rm eff}$ is larger than
$0.27$~keV$/0.22$~keV, as inferred from the temperature evolution, we
find that the energy release occurred at a depth of $\simeq 2.5$~m,
which corresponds to a column depth of $2 \times
10^{11}$g~cm$^{-2}$. This shows that the currents carrying the
magnetic field must be decaying in the upper crust. For the transient
AXP, the lack of subsequent energy release at such depths allows the
crust to cool completely, and fade back to the very low quiescent flux level.

\acknowledgements

It is a pleasure to thank Jules Halpern for suggesting XTE J1810-197
as an ideal candidate for spectral studies. We also thank Pat Slane,
Harvey Tannanbaum, Jeff McClintock, and Anatoly Spitkovsky for useful
discussions, and Dimitrios Psaltis, Ali Alpar and Fernando Camilo for
detailed comments on the manuscript. We also would like to thank the
referee Fotis Gavriil for useful suggestions. F.\"O. acknowledges
support from a Visiting Faculty fellowship from the Scientific and Technological
Research Council of TURKEY (TUBITAK). 
E.G. acknowledges partial support from the Turkish Academy
of Sciences through grant E.G/TUBA-GEB\.IP/2004-11.

\clearpage

\begin{figure}
\includegraphics[angle=-90,scale=0.5]{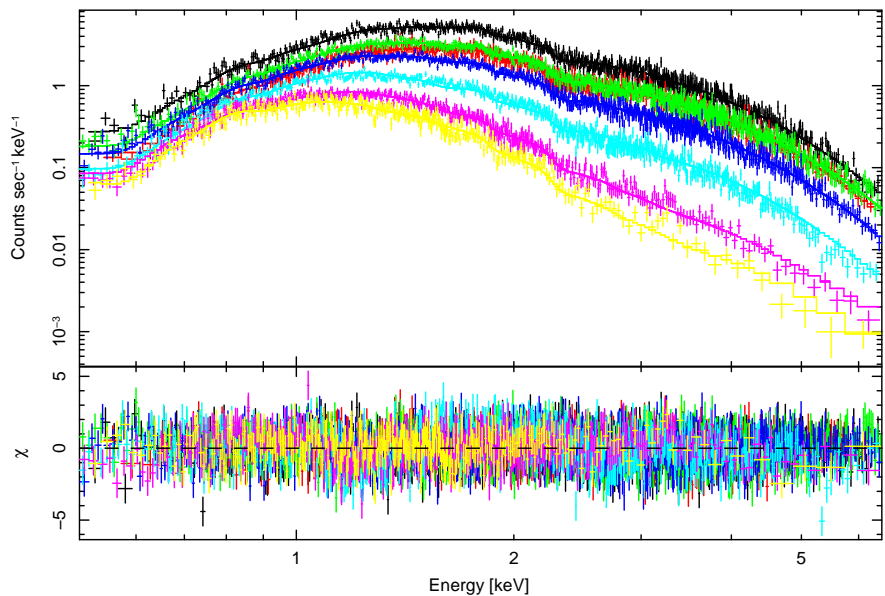}
\caption{Comparison of theoretical magnetar spectra to XMM-Newton
observations of XTE J1810$-$197 obtained over three years while the
source was declining from its 2003 outburst. Different colors
correspond to the seven different epochs of observations. Solid lines
show the best-fit theoretical models while the lower panel shows the
residuals of the fits in units of standard deviation demonstrating the
ability of the model to account in detail for the observed spectra.}
\end{figure}

\clearpage

\begin{figure}
\includegraphics[angle=0,scale=0.6]{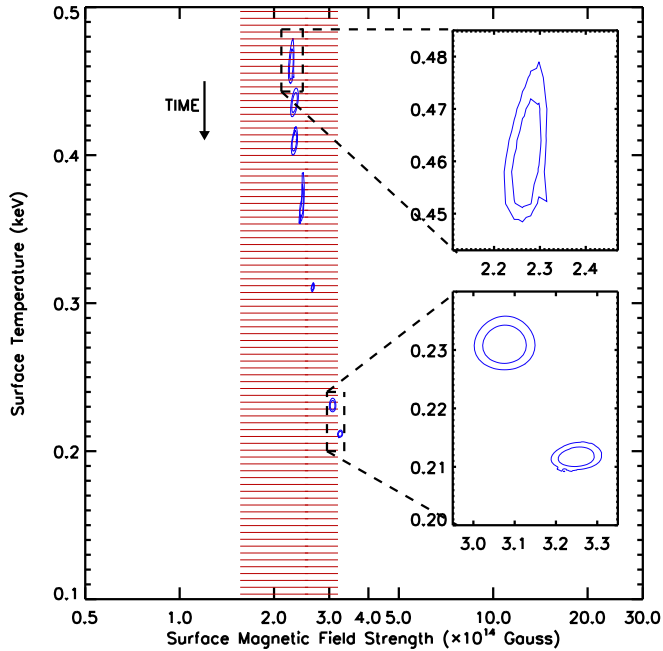}
\caption{The spectroscopically measured surface magnetic field
strength and temperature during the decline of the 2003 outburst of
XTE J1810$-$197. The contours show the one- and two-sigma confidence
limits on the parameters obtained from the individual
observations. The hatch-filled area shows the magnetic field inferred
from the observed rate of spindown, assuming magnetic dipole braking,
from the X-ray observations of Gotthelf \& Halpern (2006) and Ibrahim
et al.\ (2004), as well as the radio measurements of Camilo et al.\
(2007). The spectroscopically determined field strength is remarkably
close to the range of values inferred from the dipole spindown
formula. The monotonic and rapid decline of the measured effective
temperature is the only significant change in the source properties
during the outburst.}
\end{figure}

\clearpage

\begin{figure}
\includegraphics[scale=1.0]{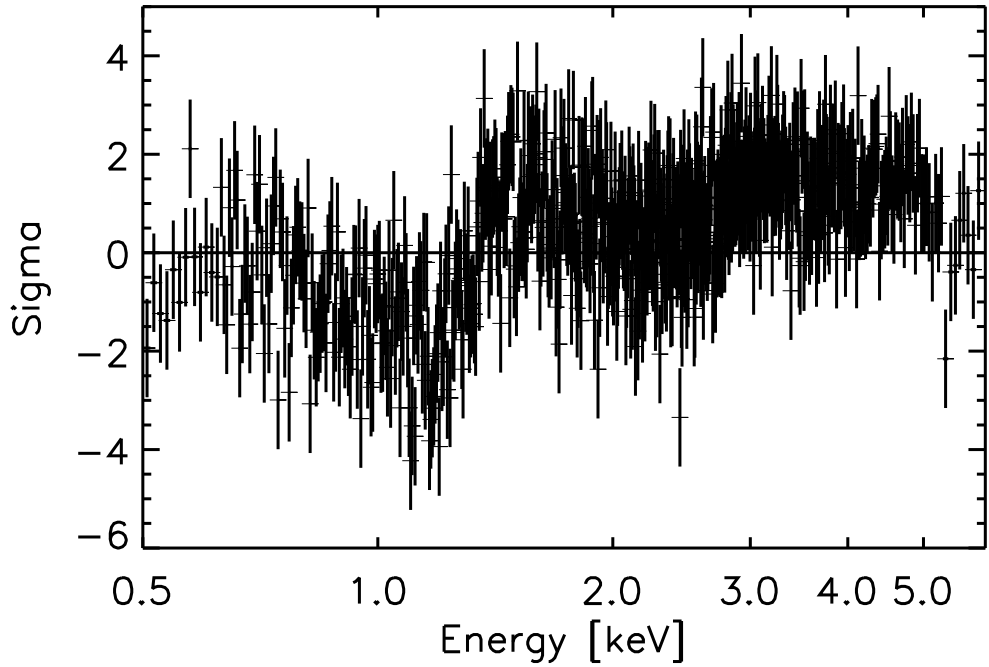}
\caption{The deviation of a test model from the data obtained on March
  18, 2005, when the magnetic field strength is artificially set
  to $2.92 \times 10^{14}$~G, that is 8-sigma higher than the best-fit
  value, while all the other parameters remain at their best-fit
  values.}
\end{figure}

\end{document}